\documentclass{article}
\usepackage{amssymb,amsmath,times}
\usepackage[pdftex]{graphics}
\usepackage{icrctc07}
\title{New method for atmospheric calibration at the Pierre Auger
Observatory using FRAM, a robotic astronomical telescope}
\shorttitle{New method for atmospheric calibration}
\authors{  Segev BenZvi, Martina Boh\'{a}\v{c}ov\'{a}, Brian Connolly,
Ji\v{r}\'{\i} Grygar, Miroslav Hrabovsk\'{y}, Tatiana
K\'{a}rov\'{a}, Du\v{s}an Mand\'{a}t, Petr
Ne\v{c}esal, Dalibor Nosek, Libor No\v{z}ka,
Miroslav Palatka, Miroslav  
Pech,  Michael Prouza, Jan \v{R}\'{\i}dk\'{y}, Petr Schov\'{a}nek, 
Radom\'{\i}r \v{S}m\'{\i}da, Petr Tr\'{a}vn\'{\i}\v{c}ek, Primo Vitale, 
and Stefan Westerhoff \\
for the Pierre Auger Collaboration$^{1}$}
\afiliations{
$^{1}$ Pierre Auger Observatory, Av. San Mart\'{\i}n Norte 304,
            5613 Malarg\"{u}e, Argentina}
\shortauthors{Pierre Auger Collaboration}
\email{travnick@fzu.cz}

\abstract{
FRAM - F/(Ph)otometric Robotic Atmospheric Monitor is the latest
addition to the atmospheric monitoring 
instruments of the Pierre Auger Observatory. An optical telescope
equipped with CCD camera and 
photometer, it automatically observes a set of selected standard stars
and a calibrated terrestrial source. 
Primarily, the wavelength dependence of the attenuation is derived and
the comparison between its vertical 
values (for stars) and horizontal values (for the terrestrial source)
is made. Further, the integral vertical 
aerosol optical depth can be obtained. A secondary program of the
instrument, the detection of optical 
counterparts of gamma-ray bursts, has already proven successful. The
hardware setup, software system, data 
taking procedures, and first analysis results are described in this paper.
}

\begin{document}
\maketitle


\section{Introduction}

FRAM is part of the Pierre Auger Observatory, and its main purpose is
to monitor continuously the atmospheric transmission.  
FRAM works as an independent, RTS2-driven \cite{rts2}, fully robotic
system, and it performs a photometric calibration  
of the sky on various UV-to-optical wavelengths using a 0.2~m
telescope and a photometer. As a primary objective, 
FRAM observes a set of chosen standard stars and a terrestrial light
source at a distance of 50\,km. From these observations it obtains instant 
extinction coefficients and the extinction wavelength dependence. The
instrument was installed during 2005 and after some
optimizations it is routinely taking data since June 2006.

The main advantage of the system is fast measurement
-- data for one star
in all filters are usually taken in less than five minutes. 
In comparison to Central Laser Facility (CLF) \cite{clf} or lidars
\cite{lidar} the other advantage of FRAM is that its measurements are 
completely non-invasive, i.e. producing no light at all and not anyhow
affecting the data acquisition of the fluorescence detector (FD). 
Furthermore,  
some of the stars are well-calibrated and non-variable standard light
sources with precisely known and tabulated  
intensities in various filters, thus allowing
straightforward comparison with measurements done using the same set
of filters. 
On the other hand, the main disadvantage of this instrument is that it
is capable only  
of integral measurements through the whole atmosphere (or for the
whole distance between telescope and terrestrial light source). 

\section{Instrument setup}

The telescope has its own laminate enclosure that is located about 30
meters from the building of fluorescence detector at Los Leones, 
on the southern edge of Auger observatory array and about 13
kilometers from the town of Malarg\"{u}e in Argentina. 

As the primary instrument we use a 20\,cm Cassegrain-type telescope
with an Optec SSP-5 photometer and with integrated 
10-position filter slider. Effective telescope focal length is 2970~mm
and focal ratio $\sim$1:15. The system is further equipped with an 
electronic focuser Optec TCF-S. This Crayford-style motorized focuser
is installed in secondary Cassegrain focus and bears the photometer  
on its moving end. A beam-splitting dichroic mirror is installed
behind the focuser. The red and infrared light is reflected into
narrow-field pointing CCD camera (Starlight XPress MX716) and
ultraviolet and visible light passes through the mirror into the
photometer.

\begin{figure}[ht]
\label{framfig}
\begin{center}
\includegraphics[width=0.4\textwidth]{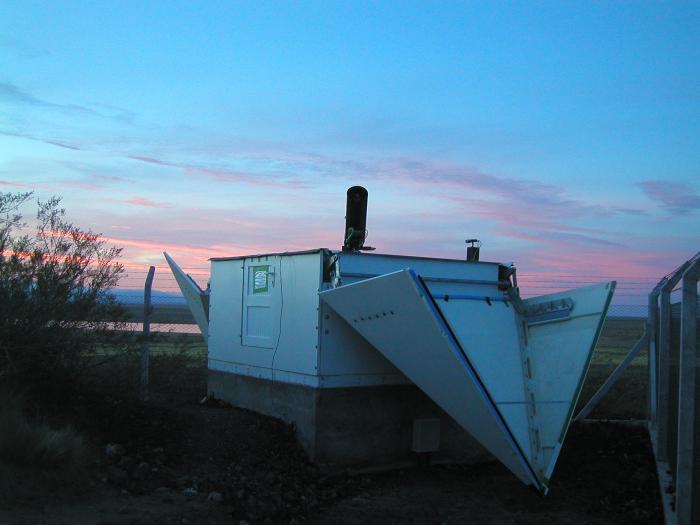}
\caption{FRAM telescope and its enclosure during sunset at Los Leones site.}
\end{center}
\end{figure}    

Narrow-field pointing CCD camera has resolution of $752 \times 580$
pixels and field of view of $7^{\prime} \times 5^{\prime}$. It is
primarily used for the fine centering of the targeted star into a
field of view of the photometer, which has only $1^{\prime}$ in
diameter. 

Photometer Optec SSP-5A is a high-precision stellar photometer. A Fabry
lens projects an image of the primary 
mirror onto the cathode of photomultiplier (PMT). The Hamamatsu R6358
PMT was selected for our setup, because of 
extended spectral response from 185 nm to 830 nm. A Fabry lens is of
B270-type glass that has enhanced UV-transmission. This still somewhat cuts 
down the transmission below 350 nm, but does not
adversely affect the transmission of any of the used filters. For star 
measurements we use the set of four Stromgren uvby filters and Johnson U 
filter, for terrestrial source observation we use also four narrowband 
filters having central wavelengths 340~nm, 365~nm, 394~nm, and 412 nm.


Atop the telescope is installed wide-field CCD camera -- Finger Lake
Instruments MaxCam CM8 with Carl Zeiss Sonnar 200mm f/2.8 telephoto
lens. This CCD camera uses Kodak KAF 1603 ME chip with $1536 \times
1024$ pixels, thus assuring $240^{\prime} \times 160^{\prime}$ field
of view. The effective diameter of the lens is 57~mm and the limiting
magnitude under optimum conditions reaches $R \sim 15.0$ for a 30~s
exposure. This CCD camera is further equipped with Finger Lake
Instruments filter wheel CFW 2 with set of Johnson-Coussins UBVRI
filters and with 380-nm and 391-nm  
narrowband filters. This wide-field CCD camera is primarily dedicated
for astrometry, i.e. for geometrical comparisons of star positions in
the images with catalogue, what then enables precise justification of
the telescope pointing on the sky.
Mount is a commercially available Losmandy G-11, which uses the
standard GEMINI GOTO 
system equipped with two servomotors with relative optical encoders.

\section{Software}

The system is driven by RTS2, or Remote Telescope System, 2nd Version,
software package \cite{rts2}.  
RTS2 is an integrated package for remote telescope control under the
Linux operating system.  
It is designed to run in fully autonomous mode, selecting targets from a
database table, storing CCD image and photometer metadata to the
database, processing images and storing their identified coordinates
in the database. RTS2 was developed and is maintained  under
open-source license in collaboration with robotic telescope projects of 
BART,
BOOTES 
and WATCHER \cite{watcher}. 

\section{Observed targets and observation schedule}

FRAM is primarily designed to provide the atmospheric extinction
model. The data
for this model are collected by the photometer with the
help of both CCD cameras. The
observation targets are selected bright (brighter than 6.5 mag) standard stars
from photometric catalogue of Perry \& Olsen \cite{olsen} that
features star measurements in Str\o mgren {\em uvby}  
photometric system.

The target cycle begins with a slew to the
position followed by a short WF camera exposure to
check the pointing accuracy. The position of the
photometer aperture within WF camera's image is well
known, so if the initial pointing is not satisfactory, a
correction could be made. This image also serves as a
test of atmospheric conditions: target may be
canceled, if the necessary conditions are not met (clouds or fog resulting in
no image astrometry).


After the star of interest was successfully centered
within the WF aperture, a control exposure with the NF CCD camera is done.
The star is identified as the brightest source in the field of view and, if needed, 
the mount position is corrected again and the star is moved 
to the center of photometer aperture.  
The photometer then does two sequences
of measurements per filter of interest. Each sequence
typically consists of five 1s integrations to obtain
the signal value and its variance in each filter.  
Simultaneously both CCD cameras take exposures, so that
pointing may be improved in real-time. The
WF camera provides also a measurement in set of Johnson-Coussins UBVRI filters.
The complete set of measurements is then stored in the structure of
PSQL database.

\subsection{Terrestrial light source observations}

The Horizontal Attenuation Monitor (HAM) \cite{ham_icrc} uses
continuously radiating Mg-Xe HID 
lamp situated about 50 km away from our telescope. We can 
observe HAM lamp also with FRAM, using both our CCD cameras and
photometer. 

The hourly automatic observations of the HAM light source started
during October 2005. The
brightest light source is identified within the 
image and then centered to the desired position on the WF camera,
using iterative procedure. When such center position is achieved, the
photometer flux is checked and observation script started.  
The ultimate goal of these observations is to obtain the
independent values for the coefficient of the aerosol attenuation
wavelength dependence and then check,  
whether the characteristic of horizontal aerosol distribution,
obtained from observations of this terrestrial source,  
agrees with the results of 'vertical' measurements of standard stars. 

\section{Optical follow-ups of gamma-ray bursts}

The RTS2 software system was originally developed 
especially for the search of optical transients
of gamma-ray bursts. This software system was significantly modified to achieve
FRAM main aims in atmospheric monitoring, however it is still very
easy to activate special observation mode for optical transients. 
The main computer of the system receives in such case the alerts
about detected gamma-ray bursts via network, slews there 
and makes images of the given sky region. 

This alert system was activated on FRAM in late 2005 and already
during January 2006 a very successful  
observation was made. An extraordinarily bright prompt optical
emission of the GRB 060117 was discovered and observed with  
a wide-field CCD camera atop the telescope FRAM from 2 to 10
minutes after GRB. Optical counterpart identified in our images
was characterized by rapid temporal  
flux decay with slope exponent $\alpha \propto 1.7 \pm 0.1$ and with a
peak brightness of 10.1 mag in Bessel R filter.  
Later observations by other instruments set a strong limit on the
optical and radio transient fluxes, unveiling an unexpectedly rapid
further decay. We presented more details in \cite{jelinek}.

\begin{figure*}[ht]
\label{framres}
\begin{center}
\includegraphics [width=0.93\textwidth]{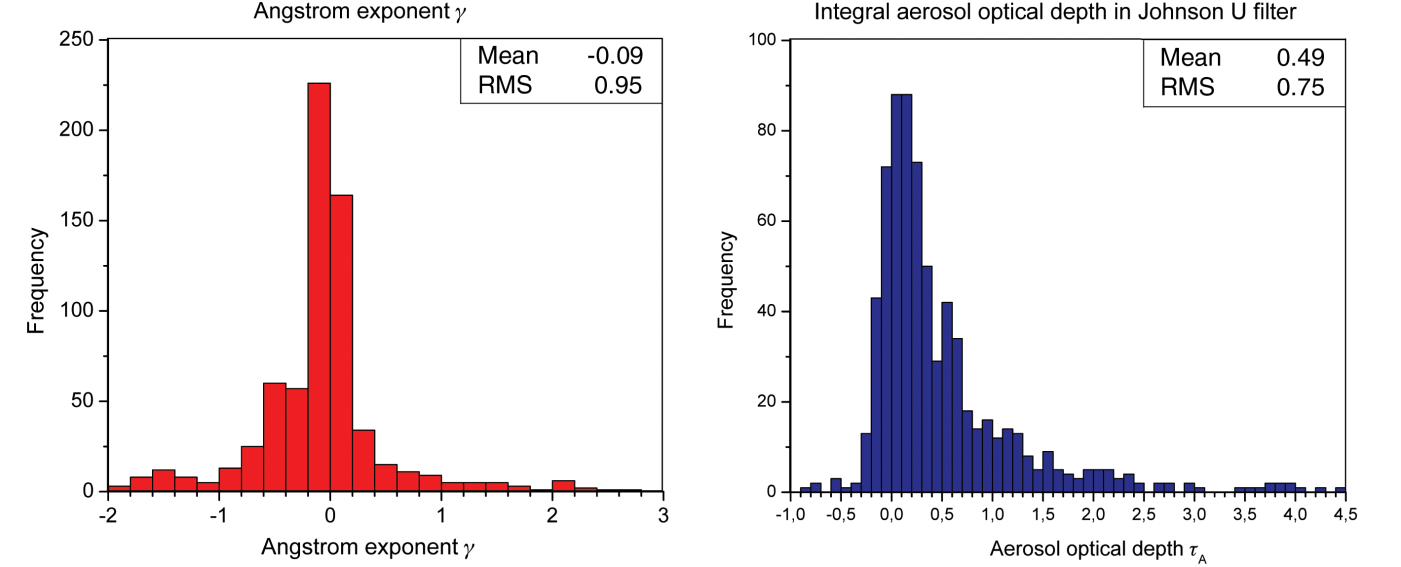}
\caption{Histograms of measured distribution of Angstrom coefficient
$\gamma$ {\it (left)} and of integral aerosol optical depth $\tau_{A}$
{\it (right)} for all data obtained from June 2006 until March
2007. Only the hardware quality cuts were applied, consequently some 
reconstruction
artifacts are still apparent. For the aerosol optical depth
distribution the left (negative) tail  is due to mis-identified 
stars and the more
prominent right tail ($\tau_{A} > 1$) is due to observations through clouds.}
\end{center}
\end{figure*}

\section{Calibration and results}

Our main goal is to provide the so-called Angstrom exponent $\gamma$,
which is often used for parametrization  
of wavelength ($\lambda$) dependence of aerosol optical depth
$\tau_{A}$: $\tau_{A} (\lambda) = \tau_{A0} \cdot (\lambda_{0} /
\lambda)^{\gamma}$, 
where $\lambda_{0}$ is the reference wavelength and $\tau_{A0}$ is the
aerosol optical depth measured for this wavelength. 
Moreover, the Johnson U filter has almost the same central
wavelength as have the lasers used for measurement of vertical aerosol
optical depth (VAOD) at 
CLF \cite{clf} and at lidar stations \cite{lidar}. 
The integral value of VAOD ($h$\,=\,$\infty$) in U filter thus can be used
for direct cross-checks with these instruments. 

We analyzed our database of photometer counts since June 2006, when
telescope entered era of stable operation, until March 2007.  
We initially used part of data for calibration, where we fitted the dependence
of difference between observed and tabulated magnitude on  
airmass for several stable and high-quality nights. The resulting 
dependence should be linear and the fit parameters characterize the
extinction (slope), but importantly also the instrument zeropoint
(intercept).  
The knowledge of the 
zeropoint then allowed us to directly compute the extinctions for our
whole database. After that, we converted extinction expressed in
magnitudes into total 
optical 
depth and then subtracted molecular Rayleigh part, using model from 
\cite{bucholtz}. 

For the standard star measurements (see Figure~2) 
we obtained a preliminary mean value of
$\gamma$\,=\,$-0.1\pm0.9$ that is lower than the results from 
HAM ($\gamma$\,=\,$0.7\pm0.5$)
\cite{aerosol}, however 
still within 1-$\sigma$ limits. Moreover, FRAM $\gamma$\,=\,$-0.1$ is
in good agreement  
with theoretical expectations for atmosphere in desert-like environment
($\gamma$\,$\sim$\,0) \cite{eck}.

In the right panel of Figure~2, a resulting peak value of VAOD 
($h$\,=\,$\infty$) is $\sim 0.15$  
for measurements in Johnson U filter. It should be noted that 
the mean VAOD value is not relevant, 
because the distribution is strongly biased with its prominent tail
of high extinction values that were obtained for observation through
clouds.  However, even the position of the peak indicates 
higher values of integral VAOD compared to the results 
from CLF and lidars (VAOD ($h$\,=\,$\infty$) $\sim$ VAOD ($h$\,=\,10~km)$ + 
0.02 = 0.08$) \cite{aerosol}. The reliable comparison will be possible
only after the application of the quality cuts 
that will follow the ongoing processing of CCD camera control images. 

{\footnotesize {\em Acknowledgments:} The telescope FRAM was built and is operated
under the support of the Czech Ministry of Education, Youth, and 
Sports through grant programs LA134 and LC527.}


\end{document}